# Weather-inspired ensemble-based probabilistic prediction of COVID-19

*Roberto Buizza*

*Scuola Superiore Sant'Anna, Pisa (Italy),*

*and Center for Climate Change studies and Sustainable Actions (3CSA; Pisa Italy)*

## Abstract

The objective of this work is to predict the spread of COVID-19 starting from observed data, using a forecast method inspired by probabilistic weather prediction systems operational today.

Results show that this method works well for China: on day 25 we could have predicted well the outcome for the next 35 days. The same method has been applied to Italy and South Korea, and forecasts for the forthcoming weeks are included in this work. For Italy, forecasts based on data collected up to today (24 March) indicate that number of observed cases could grow from the current value of 69,176, to between 101k-180k, with a 50% probability of being between 110k-135k. For South Korea, it suggests that the number of observed cases could grow from the current value of 9,018 (as of the 23rd of March), to values between 8,500 and 9,300, with a 50% probability of being between 8,700 and 8,900.

We conclude by suggesting that probabilistic disease prediction systems are possible and could be developed following key ideas and methods from weather forecasting. Having access to skilful daily updated forecasts could help taking better informed decisions on how to manage the spread of diseases such as COVID-19.

**Table of Contents**

# 1  Summary

Predicting the spread of diseases such a COVID-19 can help taking better informed decisions. The prediction should be expressed in probabilistic terms, and provide users not only with the most likely outcome but also with an objective level of confidence, which could be expressed in terms of probabilities that different scenarios could occur.

An ensemble of 30 members, with parameters estimated by randomly perturbing the observed data in a way to simulate observation errors, is shown to provide valuable forecasts of the most likely outcome, possible future ranges and probabilities.

Results show that:
- For the case of China, this method worked well, and on day 25 could have predicted the outcome for the next 35 days;
- For the case of Italy, forecasts based on data collected up to the time of writing (24 March) indicate that number of observed cases could grow from the current value of 69,176, to between 101k-180k, with a 50% probability of being between 110k-135k;
- For South Korea, forecasts suggest that the number of observed cases could grow from the current value of 9,018 (as of the 23rd of March), to values between 8,500 and 9,300, with a 50% probability of being between 8,700 and 8,900.

It is suggested that methods developed from ensemble-based weather forecasting could be followed to design and develop ensemble-based probabilistic diseases prediction systems, so that decision makers can take better-informed decisions.

# 2  COVID-19: infection data analysis and prediction

Since the initial spread of the COVID-19 infection, many methods have been proposed and applied to try to predict future numbers, either based only on data analysis, or on health models. This work falls into the first category. More precisely, it proposes a method inspired by the ensemble prediction systems developed in the past 25 years to predict the weather [1, 2].

If we look at Italy, which today is facing a very critical situation, some authors (see e.g. [3]) have spoken of an 'exponential growth rate'. De Nicolao, for example, on 4 March talked about the possibility that numbers would reach 700+. Unfortunately, he severely underestimated the numbers to be reached in the following days of the 'total assessed cases', which the Italian Civil Protection (PC) site has reported to have reached 17,660 by 18.00 of 13 March (at the time of writing).

Other authors (see, e.g., [4]) pointed out that the growth rate of diseases is not exponential, but follows the 'S-shape' curve described by a logistic equation. [5] has also analysed the Italian data and tried to fit the two curves to the existing data up to 12$^{th}$ March, and concluded that it is still very difficult to predict the future number of assessed cases.

Compared to an exponential curve, a logistic curve has an asymptotic value, and thus cannot grow to an infinite value as the exponential curve does. This growth limitation is partly due to the fact that there is a physical limit to the number of people who can be infected, and partly due to measures that can be implemented to contain the spread of the disease. As soon as the containment measures start working, one should expect that the growth rate slows down, the curve changes concavity, and starts evolving toward a curve that resembles more a limited one with an asymptotic value. The solution of a logistic equation behaves precisely in this way. Indeed, a logistic model has been used, for example, to predict the risk of developing a given disease (e.g. diabetes; coronary heart disease), based on observed characteristics of the patient [6, 7].

In particular, the following three questions are going to addressed:
a) Is the logistic equation capable to fit well observed data of COVID-19?
b) Can we develop diseases' prediction methods following the example of ensemble-based probabilistic weather prediction?
c) Can we use an ensemble of stochastically-perturbed logistic curves to generate COVID-19 probabilistic predictions of future infection numbers?

In section 2, the logistic equation is introduced, and an ensemble-based Monte Carlo system designed to estimate future forecast probabilities is described. In section 3, COVID-19 data from China are analysed using the logistic equation, and tests are performed to assess whether and when (i.e. how many days after the first case has been reported) the numbers of the subsequent days can be predicted. In section 4, the same analysis is be applied to Italy and South Korea. Conclusions are then drawn in Section 5.

## 3   The growth model

The growth model equation used to analyse COVID-19 infection data, and predict future numbers, is based on the same equation used to investigate and understand the predictability of weather systems. The equation, first proposed by [8] *Lorenz* (1982), then modified by [9, 10], was used by [11] to investigate how resolution increases could improve weather forecasts.

The fact that a model characterized by sub-exponential growth rates, rather than an exponential one, could fit better the data was pointed out also by authors studying the propagation of COVID-19. For example, [12] stated that *'.. The recent outbreak of*

*COVID-19 in Mainland China is characterized by a distinctive algebraic, sub- exponential increase of confirmed cases with time during the early phase of the epidemic, contrasting an initial exponential growth expected for an unconstrained outbreak with sufficiently large reproduction rate.'.*

Our choice for a logistic curve was inspired by its use in weather prediction, as both problems face a maximum asymptotic value, and growth rates are dominated by linear and quadratic terms [8, 11]. Indeed, the equation which has as solution a logistic curve includes both a linear and a non-linear term:

(1a) $\quad \frac{dE}{dt} = \propto E - \beta E^2 + \gamma$

where $E$ is a measure of a quantity that is growing: the forecast error in weather prediction, or, in this work, the cumulative number of infected cases.

Eq. (1a) can also be written as:

(1b) $\quad \frac{dE}{dt} = (\propto E + S)\left(1 - \frac{E}{E_\infty}\right)$

where

(2) $\quad \begin{cases} S \equiv \gamma \\ E_\infty \equiv \frac{\alpha}{2\beta} + \sqrt{\frac{\alpha^2}{4\beta^2} + \frac{\gamma}{\beta}} \\ a \equiv \beta E_\infty \end{cases}$

The solution of Eq. (1) is given by:

(3a) $\quad E(t) = E_\infty \cdot \left[1 - \frac{1}{1+C_1 e^{C_2 t}}\left(1 + \frac{S}{aE_\infty}\right)\right]$

where $E_\infty$ is the asymptotic value. Eq. (4a) can be written also in normalized form as:

(3b) $\quad \eta(t) \equiv \frac{E(t)}{E_\infty} = 1 - \frac{1}{1+C_1 e^{C_2 t}}\left(1 + \frac{S}{aE_\infty}\right)$

The solution coefficients $C_1$ and $C_2$ are given by:

(4) $\quad \begin{cases} C_1 \equiv \frac{1}{E_\infty - E_0}\left(E_0 + \frac{S}{a}\right) \\ C_2 \equiv a + \frac{S}{E_\infty} \end{cases}$

where $E_0$ is the initial-time error.

As [9] indicated, when used in weather prediction we can interpret *a* as the initial rate of growth of E, *S* as the effect of model uncertainties on the error growth, and $E_\infty$ as the error asymptotic value. The (α,β,γ) parameters of Eq. (1a) can be determined by fitting the curve to the data. Once these parameters have been computed, the coefficients *a, S* and $E_\infty$ can be determined by applying the definitions (4), and the analytical solutions (3a,b) can be computed (see, e.g., [11]).

The solutions (3a,b) can be used to predict future values. Eq. (3b) can also be used to compute the doubling time at each time step t:

(5) $\quad \tau(t) = \frac{\ln(2)}{\alpha + \frac{\gamma}{E(t)}}$

# 4   An ensemble-based Monte-Carlo system to estimate confidence and future scenarii

The main weakness of generating forecasts using only a single, deterministic forecast is that it does not provide any confidence information, for example expressed in terms of the possible range of future values (see, e.g., discussions in the articles collected in [1]. Furthermore, it does not take into account the fact that the training data used to estimate the model' parameters are affected by observation errors, and this can lead to forecast errors. In other words, it does not take into account initial condition's uncertainties. The analogous in weather prediction is the generation of an ensemble of initial states for the atmosphere or the ocean, by assimilating perturbed observations.

We could address these two weaknesses by using an ensemble of forecasts, based on logistic curves estimated by perturbing the training data by an amount that reflects the observations' uncertainty. We can then use the ensemble of forecasts to estimate the future range of scenario, which could be expressed, for example, in terms of the minimum and maximum value, and quantiles computed from the predicted ensemble members. Similar approaches are followed in weather prediction since 1992: see, e.g., [1, 2] for an overview.

This is precisely the method that we have followed: observed data covering a training period (e.g. data up to today) have been stochastically perturbed, and an ensemble of 30 forecasts have been generated. Each forecast is been defined by a logistic curve, with its governing parameters $(\alpha_j, \beta_j, \gamma_j)$, with j=1, ..33, estimated by applying best-linear fit methods to the perturbed observations.

Each observation has been perturbed by a random number, sampled from a Gaussian distribution with a standard deviation defined to be 10% of the daily variations in the *Roberto Buizza – Ensemble prediction of COVID-19 – BMJ Open (24 March 2020)*observed value. By stochastically perturbing the observations used to compute the parameters of each logistic curve, we generate an ensemble of curves, each defined by a slightly different parameters. In this way, we simulate not only initial condition errors, but also model errors.

To further improve the simulation of initial condition errors, and to take into account the fact that data might be affected by a timing error, and thus could be over or under estimated, 10 of the 30 members are shifted forward by one day, and 10 backward by one day.

Suppose, for example, that it is the evening of day 31 since the start of an infection (23$^{rd}$ of March 2020), and we have access to observations of the past 31 days since the number has jumped from zero to a positive value. This is how the forecasts for the future days are generated:
- The daily percentage increment in the numbers $\delta n_{obs}(d)$ is computed for each of the past 31 days;
- The standard deviation $\sigma_{obs}$ of the increments is computed;
- 30 sets of observations are defined by randomly perturbing each original (unperturbed) observation by a random number *r(d)* sampled from a Gaussian distribution with zero mean, and standard deviation given by $\sigma_{obs}$; each perturbed observation is defined by multiplying the observed value by *[1+r(d)]*, and 30 logistic curves are fitted to the data; these 30 ensemble members are defined by logistic curves (as in Eq. 3a) with slightly different parameters;
- To take into account
- Statistics (e.g. the ensemble-mean and the ensemble standard deviation) are computed using the 30 curves, to estimate the probability distribution of future numbers.

The idea behind this approach is to use the statistics of the daily changes in the observed numbers as an approximation of the statistic of possible observation errors: given no other indications of the possible observation errors, we believe that this could be a reasonable choice. The idea of perturbing input observations has been inspired by the successful use of similar approaches in weather prediction (see, e.g., [13, 14, 15]). A similar approach is also used in ensembles generated by lagged forecasts, i.e. forecasts with different initial states, whereby it is assumed that one could simulated initial condition uncertainties by using information (in this case the analysis) of consecutive days.

We recognize that the choice of sampling from a Gaussian distribution with a standard deviation equal to 0.1 times $\sigma_{obs}$ is an arbitrary choice that should be further tested in the future. Sensitivity tests for China (see Section 4) suggest that this is a reasonable choice, and for the purpose of this work is reasonable and can help us illustrating the value of an ensemble-based, probabilistic approach compared to one based on a single prediction.

## 5   COVID-19: the case of China

The left panel of Fig. 1 shows the cumulative observed number of confirmed cases from China, reported from the 22$^{nd}$ of January to the 7$^{th}$ of March, and a logistic curves with parameters estimated using all the observed numbers. It shows that the logistic curve is capable to describe very well the data: the asymptotic value of the logistic curve is 68,790, and the cumulative number of cases reported on the 7$^{th}$ of March is 68,482 (see Table A.1 in the Appendix for the WHO data used in this study). It is worth pointing out that on 13 February, WHO reported a value of 15,200, while the day before and the day after it reported 2,000 and 4,000: because of the very large difference between 15,200 and these two values, we decided to replace 15,200 with the average between the values reported on the 12$^{th}$ and the 14$^{th}$ of February. A similar approach is used in weather prediction, when quality-control methods are applied to ensure that the data used to initialize the models are not affected by unexpected, large errors.

Figure 1 shows the comparison of the observed data, with a logistic curve estimated using only data from the first 25 days. Note that although the fit is not as accurate as when all 45 days of data are used, there is a very high-correlation of 99.8% for forecasts issued on day 25, with a root-mean-square error of 1,603, compared to 1,196 for a logistic curve generated assimilating observations of all 45 days.

Two very interesting questions to ask are the following:
- Since when could we have predicted correctly the asymptotic value?
- Could we identify when this would have been possible by looking at the observed data?

Figure 2 shows the data from observed data from China during the first 30 days, Figure 3 shows its first order derivative and Figure 4 its second order derivatives in time. The first order derivative gives us an information on the growth rate, and the second order derivative tells us whether the data follow a convex or a concave trend. At around day 15 the derivative starts decreasing in amplitude (Fig. 3), and the second derivative becomes negative (Fig. 4), indicating that the concavity of the curve changes. The fitted curves (the dotted lines), clearly, show smoother changes. Day 18, when the second derivative changes sign, is when the fitted curve (Fig. 1) starts bending towards an asymptotic value, and stops behaving as an exponential curve. The problem with the predictions for China is that numbers started climbing again between day 21 and 23: during these days, the first derivatives increases and the concavity of the curve changes, again, and becomes positive.

By looking at the growth rate and the concavity of the observations curve, we can identify whether the observed curve starts behaving like a logistic curve, and thus we could try to predict the range of possible values using our ensemble approach, based on stochastically-perturbed logistic curves.

Going back to the two questions that we posed above, we can then say that:
- We cannot aim to predict the asymptotic value until we do not detect that the observed data show a clear transition from an exponential growth to one similar to a logistic one;
- We could identify when this would be possible by analysis the observed data, and in particular at its first and second order derivatives.

In other words, the simple, error-growth model has some limitations in its predictions' capabilities. As it is the case for weather prediction, simple models have the advantage that it is easier to interpret their behaviour than complex models, but they have limited predictive skill.

## 6   COVID-19: ensemble-based predictions for China

We should try now to predict the number of COVID-19 total cases, starting from days before and after day 25 from the start of the infection. The data shown in Figs. 3-4 indicate that only after day 24 there was a clear indication that numbers were growing at a lower rate, and the curve's concavity was turning negative. Thus, we should expect difficulties in predicting the future numbers before day 25.

Figures 5 and 6 show two probabilistic forecasts, generated on day 22 and day 25. Each probabilistic forecast is expressed in terms of five curves: the minimum and maximum values, the median, and the $25^{th}$ and $75^{th}$ quantiles (no assumption has been made on the shape of the forecast probability density function).

Although one cannot judge the value of a probabilistic prediction system on a single case, we could see that the observed values are outside the range spanned by the forecast distribution generated on day 22 (Fig. 5), but were included in the range predicted on day 25 (Fig. 6).

If we consider the forecast issued at day-25, results indicate that:
- There was a ~25% probability that the number would be lower than the 60,000 (i.e. below the $25^{th}$ quantile);
- There was a ~25% probability that numbers would be higher than 75,000 (i.e. above the $75^{th}$ quantile);
- The most-likely outcome, given by the median forecast, gave for day 45 a number not far from the observed value;

- The observed values between day 25 and 45 were inside the range spanned by the forecast probability distribution.

# 7  COVID-19: ensemble-based predictions for Italy and South Korea

At the time of writing this document (24 March 2020), the situation in Italy is very serious. Today is the 32$^{nd}$ day since the first case was reported, and from the site of the Italian Protection Agency one could have access to 32 days of confirmed number of cases (see Table A.2, in the Appendix). Authorities are struggling to estimate how fast, and which number of cases they could be facing in the next few weeks. South Korea and the UK have also been facing an increasing number of cases.

Can we say something about the future numbers of total number of infected cases, starting from the observed data, in Italy, South Korea and the United Kingdom and?

Figure 7 shows the same observed curves shown in Fig. 2 for China, and the corresponding observed curves for Italy, South Korea and the UK (for all countries, day 1 is the first day when infected cases were reported). Figures 8 and 9 show the first and second order derivatives, i.e. the daily increments of the total number, and the concavity of the observed curve. If we look at the first and second order derivatives, we can see that:
- *Italy*: values are still rising fast and have reached China; the first derivative indicates that the last two days have shown a slower increase, and the second derivative indicates that concavity has become negative;
- *UK*: the first derivative is still increasing, and the concavity is still positive;
- *South Korea*: the first derivative have been showing a slower increase than the other three countries, and the concavity of the curve has been very close to zero, with days with a slightly positive value followed by days with slightly negative values.

If we apply the reasoning that helped us understanding forecasts for China, reported in Section 3, we can deduce that for both Italy and South Korea, it is still impossible to predict the range of possible values that could be reached in Italy in the next 30 days. Indeed, for both countries our minimisation algorithm has not bene able to fit logistic curves to the observed values. We need to wait until we can detect a clear slowdown in the growth rate and a clear change in the concavity of the curve, before we can predict the possible future range that will be reached asymptotically, as it was the case for China.

Figures 8-9 indicates that for Italy, and possibly South Korea, we should be able to generate probabilistic forecasts.

Figures 10-11-12 shows the forecasts for Italy, issued on day issued on day 30 (22 March), day 31 (23 March) and day 32 (24 March, at the time of writing this paper). Forecasts could not be generated in the days before day 30 because the total number of cases was still increasing without any detectable change in the curve concavity, and the minimisation algorithm was not able to find logistic curves that would fit the observed date. The three consecutive forecasts are rather consistent: they all predict that in 30 days from now the total number of cases would be between 101k-180k, with a 50% chance of being between 110k-135k. Note that there is still a rather large uncertainty, especially in the older two forecasts.

Figure 13-14-15 shows the forecasts for South Korea, issued on day 20, 30 and 33 (24 March, with data up to 23$^{rd}$ March). Note that, compared to the Italian case, there is less uncertainty in the forecasts: this is a consequence of the fact that data has been more stable, growing at a slower pace than Italy. Note that the earlier forecast, issued on day 20, did not include the observations collected during days 21-33. As can be deduced from Figs. 8-9, this is due to the fact that, during these days, the numbers for Korea have started increasing again (positive first derivative) and the curve has shown a small but positive concavity. This might be due to new clusters of infection having started in different part of Korea. The fact that our model cannot predict this is due to its coarse resolution, to the fact that it aggregates the data of a whole country without distinguishing them.

Similar results occur in weather prediction, when one tries to predict detailed weather patterns with a model with too coarse a resolution, either in physical space or in probability space. Here, we suffer from the same problem: more complex models capable to simulate different regions, or clusters, separately, should be used if one wants to be able to resolve these small scale features.

## 8   Conclusions

If this work we discussed issues linked to the prediction of the evolution of the COVID-19 disease, and we proposed an ensemble-based, probabilistic approach inspired by ensemble methods used in operations in weather prediction. The key advantage of following a probabilistic approach is that it allows to take into consideration initial condition and model uncertainties, and it allows to predict the future range of possible scenarii.

In particular, we investigated whether we were able to predict the future number of COVID-19 (see, e.g., [16]) infections for a country such as China, Italy, South Korea and the United Kingdom, using an ensemble of stochastically-perturbed logistic curves. Each single curve has been obtained by perturbing the initial conditions, to simulate initial uncertainty, and by perturbing the logistic curve parameters, to simulate model uncertainty.

The first result of this work is that, although the model is very simple, it allows to generate realistic probabilistic forecasts. Realistic in the sense that the curve fitted well the observed data for China, and provides realistic forecasts for Italy and South Korea, capable to include the observations, for the few cases to which it has been applied. Although we cannot make stronger, more robust conclusions, since the method has been tested on few cases, we believe that this work could inspire future developments by providing some key ideas, which have been tested thoroughly in weather prediction.

Results based on the most recent data collected today (24$^{th}$ March 2020), indicate that it is still too early to make predictions for the United Kingdom, because the observations do not show yet that the growth rate has slowed down, and the concavity of the observed curve has yet changed sign and become negative. For Italy and South Korea, the ensemble method provides forecasts for the next 30 days:
- For Italy, it suggests that the number of observed cases could grow from the current value of 69,176 (as of today, 24$_{th}$ of March), to values between 101k-180k, with a 50% probability of being between 110k-135k;
- For South Korea, it suggests that the number of observed cases could grow from the current value of 9,018 (as of the 23$_{rd}$ of March), to values between 8,500 and 9,300, with a 50% probability of being between 8,700 and 8,900.

This ensemble-based approach, inspired by works in the field on numerical weather prediction is, to our knowledge, novel for this field. The ensemble curves were generated by simulating initial and model uncertainties using stochastic perturbations. Similar ensemble-generation approaches, whereby observations and models are stochastically perturbed to generate the ensemble of single forecasts, have been used in operational weather prediction very successfully since 1992 [1, 2, 13, 14, 15].

The choice of the logistic equation as the basis for each single forecast calculation is based on the fact that sub-exponential growth rates, rather than exponential ones, would fit better observed data of the spread of diseases, as was pointed out by other authors. For example, [12] stated that *'.. The recent outbreak of COVID-19 in Mainland China is characterized by a distinctive algebraic, sub- exponential increase of confirmed cases with time during the early phase of the epidemic, contrasting an initial exponential growth expected for an unconstrained outbreak with sufficiently large reproduction rate.'*. Similar considerations dictated such a choice in other fields, for example in predictability studies of the weather, since also in this case there is a maximum asymptotic value, and growth rates are dominated by a combination of linear and quadratic terms [8, 11].

Two key questions have been addressed in this study:
a) Is the logistic equation capable to fit well observed data of COVID-19?
b) Can we develop diseases' prediction methods following the example of ensemble-based probabilistic weather prediction?
c) Can we use an ensemble of stochastically-perturbed logistic curves to generate COVID-19 probabilistic predictions of future infection numbers?

Results indicate that:
a) The logistic curve can fit reasonably well the data of COVID-19 spread;
b) We can develop ensemble-based prediction methods inspired by weather ensembles;
c) We can use this equation to generate probabilistic predictions, but we can do this only from the time when we the observed data show a clear transition from an exponential growth to one similar to a logistic one; this time could be identified by analysing the observed data, and in particular their first and second order derivatives.

We hope that this work can promote the development of more realistic, accurate and reliable ensemble methods capable to predict the spread of diseases such as COVID-19, and that in the future we will see, as it is the case for weather, ensemble-based probabilistic predictions being at the core of disease monitoring and risk management.

Ensemble-based probabilistic methods have been used for almost three decades to predict possible future weather scenarii. We suggest that weather-inspired ensemble-based probabilistic main concepts, ideas and methods could be used applied to predict COVID-19 spread. It is shown that valuable probabilistic forecasts can be generated using an ensemble of stochastically-perturbed logistic curves, as the ones first used by [8], then modified by [9], and thereafter used by many other authors to investigate forecast error growth. Ensemble members are generated by perturbing the input observations, following approaches used to generate atmospheric and ocean ensembles of analyses.

## 10 References

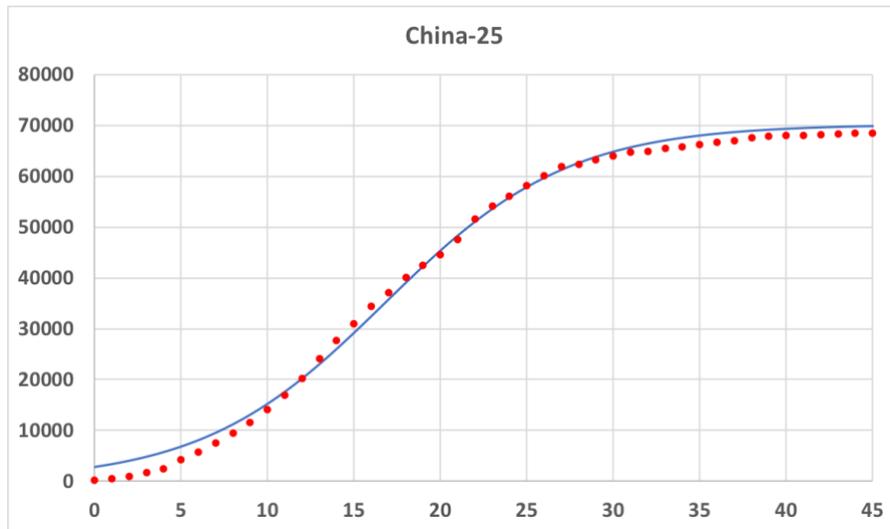

*Figure 1. COVID-19 China reported cases (red dots; values from WHO), and fitted logistic curve (blue lines), with parameters estimated using data covering different periods. The logistic curve has parameters estimated using data covering only the first 15 days.*

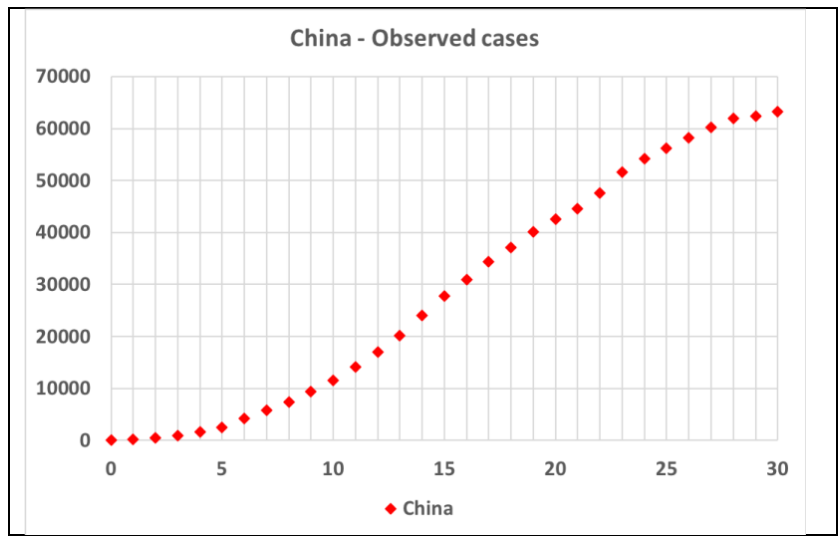

*Figure 2. COVID-19 China. Reported cases during the first 30 days (day 1 is the first day when a number different from zero was reported; data from WHO).*

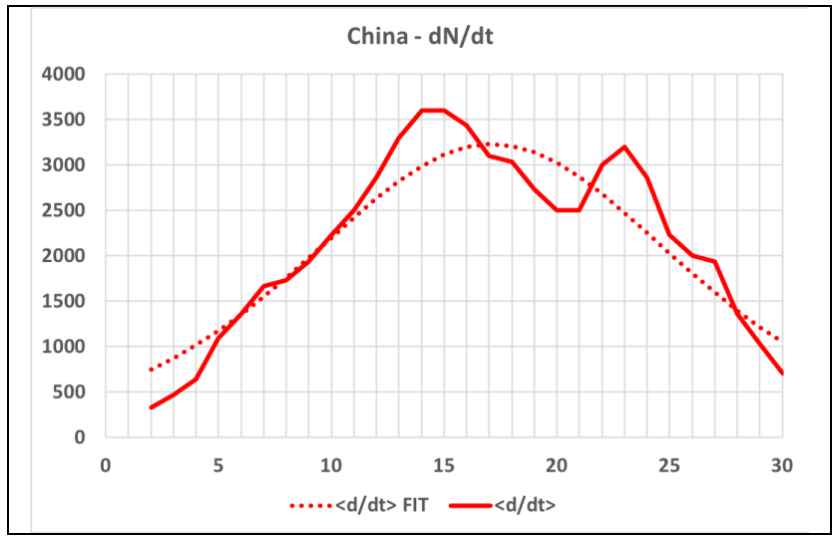

*Figure 3. COVID-19 China. Observed daily increments (i.e. the derivative of the observed curve shown in Fig. 2; solid line) and increments computed from the fitted curve (dotted line).*

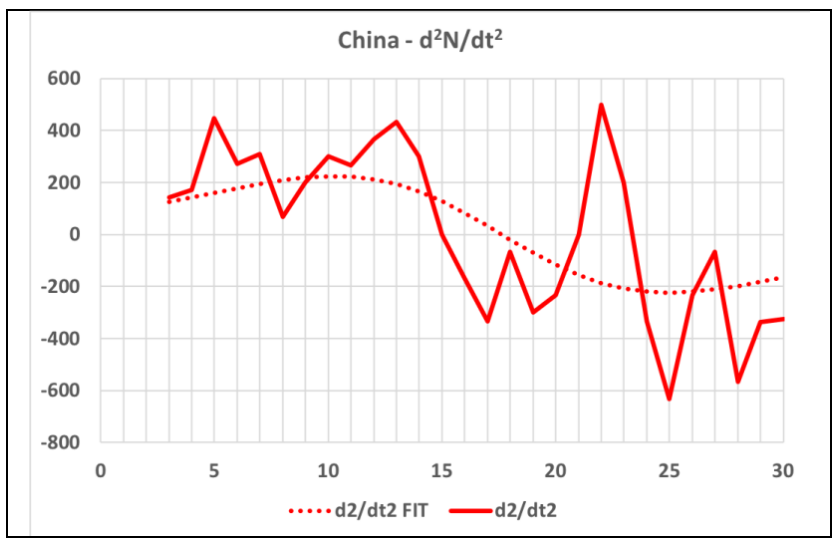

*Figure 4. COVID-19 China. Second derivative of the observed daily increment curve (solid line) and second derivative of the fitted curve (dotted line).*

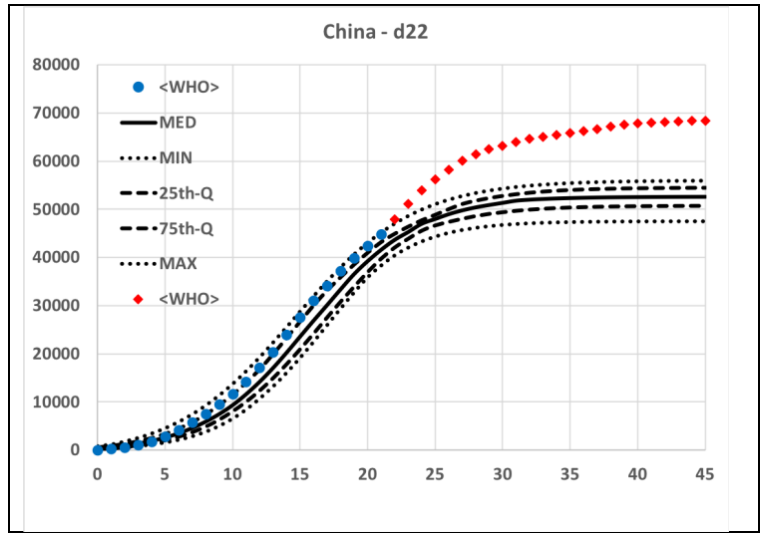

*Figure 5. COVID-19 China. Ensemble-based probabilistic forecast issued on day 22. The blue symbols show the observations used to estimate the ensemble of 30 logistic curves. The red symbols show the reported number of confirmed cases of days 23-to-45. Five black curves are shown: the minimum and maximum values (dotted lines), the median (solid line), and the 25$^{th}$ and 75$^{th}$ quantiles (dashed lines).*

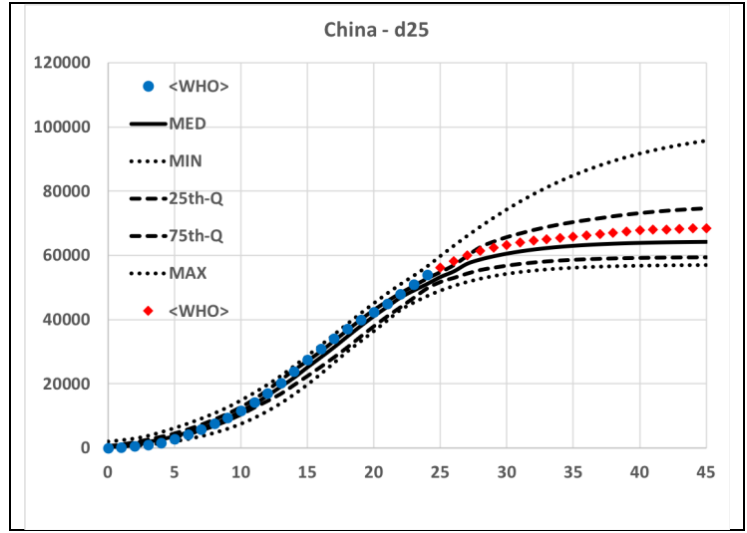

*Figure 6. COVID-19 China. Ensemble-based probabilistic forecast issued on day 25. The blue symbols show the observations used to estimate the ensemble of 30 logistic curves. The red symbols show the reported number of confirmed cases of days 26-to-45. Five black curves are shown: the minimum and maximum values (dotted lines), the median (solid line), and the 25$^{th}$ and 75$^{th}$ quantiles (dashed lines).*

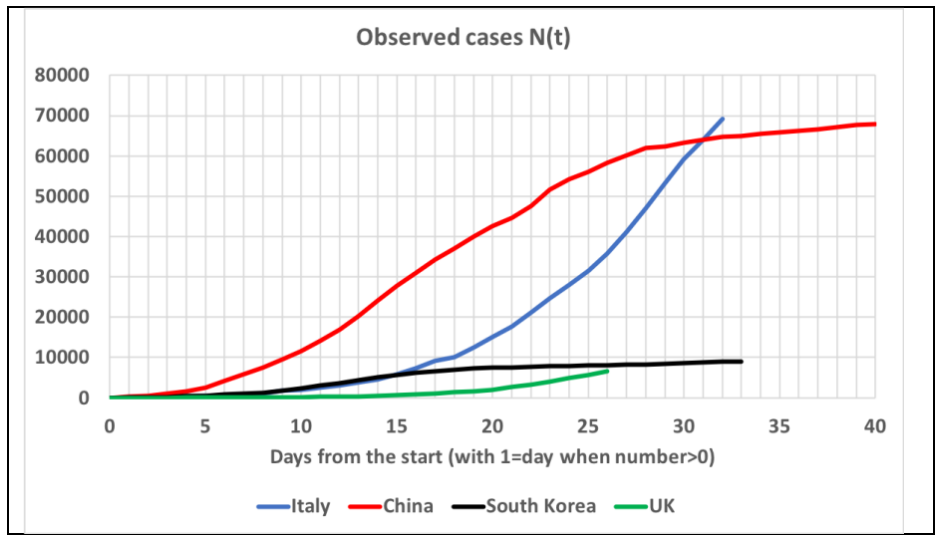

*Figure 7. COVID-19 China (red lines), Italy (blue lines), the UK (green lines) and South Korea (black lines). Total number of reported (day 1 is the first day when a number different from zero was reported; data from WHO for China, the UK and South Korea, and the Civil Protection Agency for Italy).*

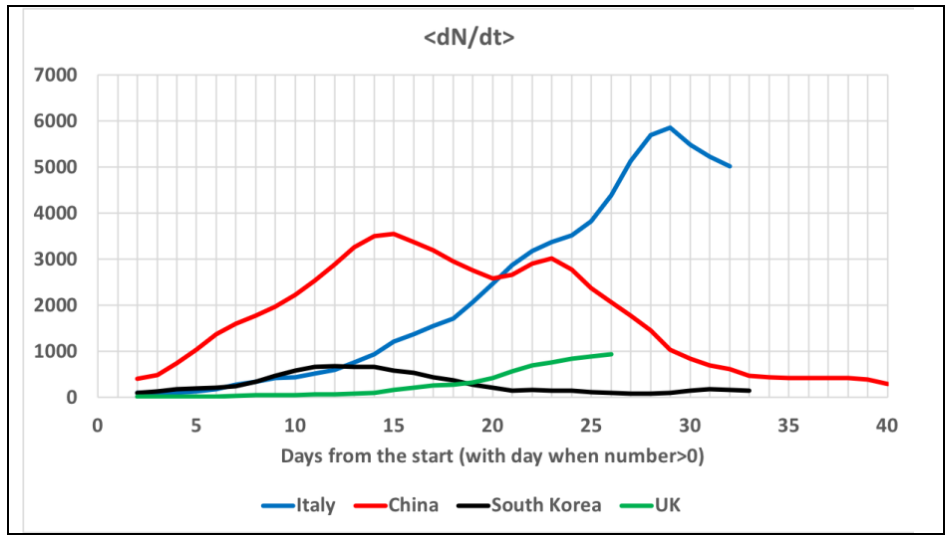

*Figure 8. COVID-19 China (red lines), Italy (blue lines), the UK (green lines) and South Korea (black lines). Observed daily increments, averaged over a centred 3-day interval (i.e. the derivative of the observed curve; each day, centred 3-day running mean averages are shown).*

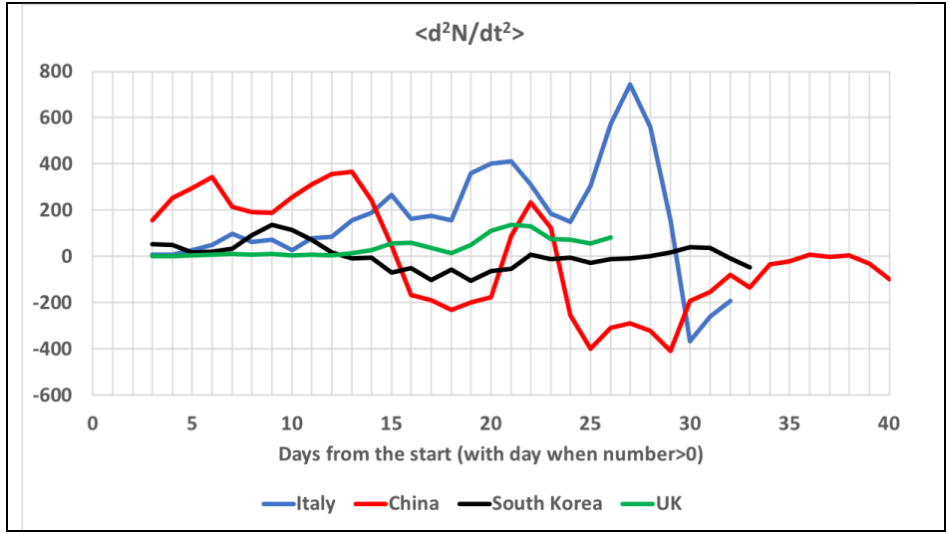

*Figure 9. COVID-19 China (red lines), Italy (blue lines), the UK (green lines) and South Korea (black lines). Concavity (i.e. the second derivative) of the observed curve, averaged over a centred 3-day interval.*

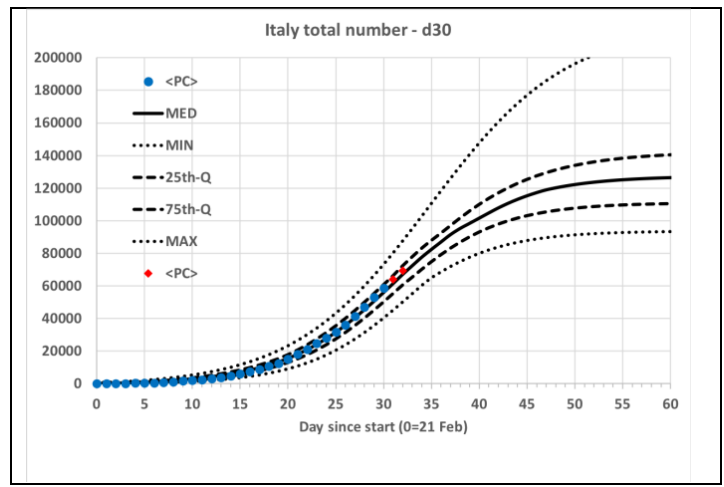

*Figure 10. Forecasts of the 'reported number of cases' for Italy, issued on day 30 (22 March). The blue symbols show the observations (from WHO) used to estimate the ensemble of 30 logistic curves. The red symbols show the observations not used to compute the curves, covering days 31-32 (at the day when the computation was completed, 24 March). Five black curves are shown: the minimum and maximum values (dotted lines), the median (solid line), and the 25$^{th}$ and 75$^{th}$ quantiles (dashed lines).*

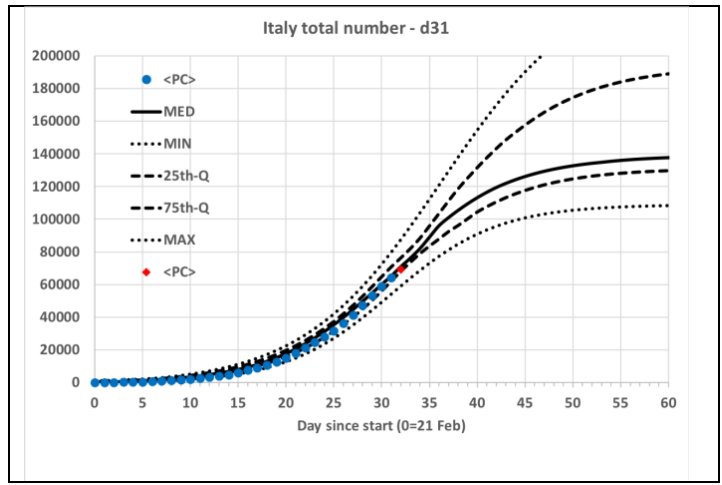

*Figure 11. Forecasts of the 'reported number of cases' for Italy, issued on day 31 (23 March).The blue symbols show the observations (from WHO) used to estimate the ensemble of 30 logistic curves. The red symbols show the observations not used to compute the curves, covering days 32 (at the day when the computation was completed, 24 March). Five black curves are shown: the minimum and maximum values (dotted lines), the median (solid line), and the 25$^{th}$ and 75$^{th}$ quantiles (dashed lines).*

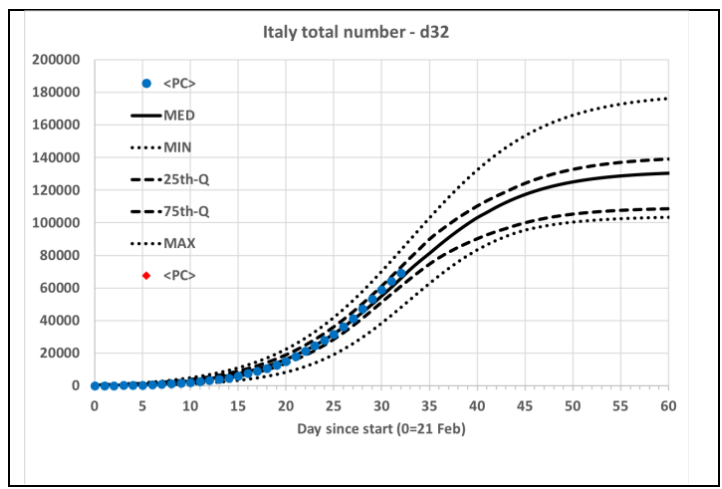

*Figure 12. Forecasts of the 'reported number of cases' for Italy, issued on day 32 (24 March).The blue symbols show the observations (from WHO) used to estimate the ensemble of 30 logistic curves. Five black curves are shown: the minimum and maximum values (dotted lines), the median (solid line), and the 25$^{th}$ and 75$^{th}$ quantiles (dashed lines).*

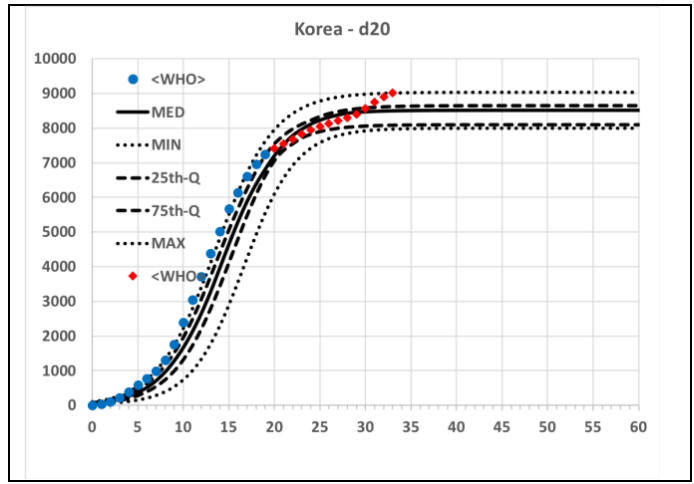

*Figure 13. Forecasts of the 'reported number of cases' for Korea, issued on day 20 (8 March). The blue symbols show the observations (from WHO) used to estimate the ensemble of 30 logistic curves. The red symbols show the observations not used to compute the curves, covering days 21-33. Five black curves are shown: the minimum and maximum values (dotted lines), the median (solid line), and the $25^{th}$ and $75^{th}$ quantiles (dashed lines).*

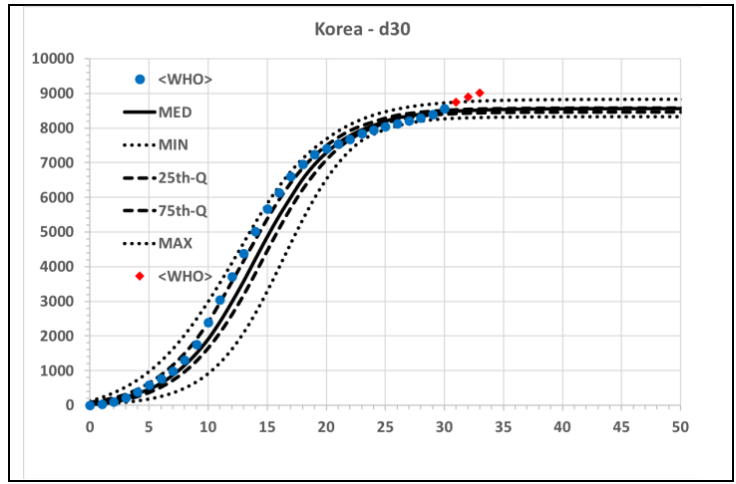

*Figure 14. Forecasts of the 'reported number of cases' for Korea, issued on day 30 (19 March). The blue symbols show the observations (from WHO) used to estimate the ensemble of 30 logistic curves. The red symbols show the observations not used to compute the curves, covering days 31-33. Five black curves are shown: the minimum and maximum values (dotted lines), the median (solid line), and the $25^{th}$ and $75^{th}$ quantiles (dashed lines).*

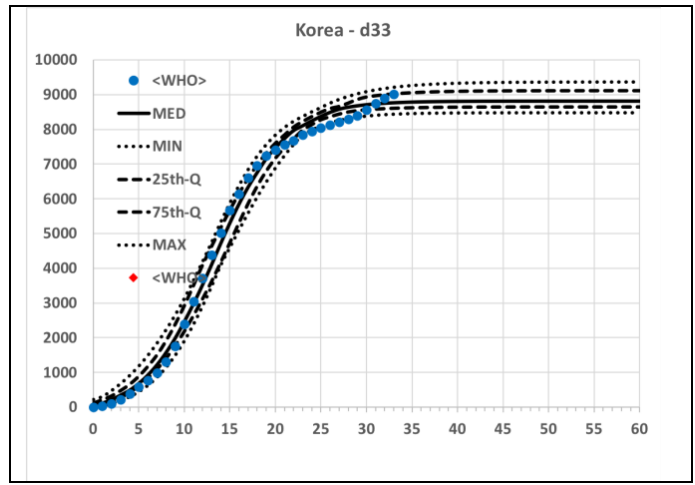

*Figure 15. Forecasts of the 'reported number of cases' for Korea, issued on day 33 (22 March).The blue symbols show the observations (from WHO) used to estimate the ensemble of 30 logistic curves. Five black curves are shown: the minimum and maximum values (dotted lines), the median (solid line), and the 25th and 75th quantiles (dashed lines).*

*Roberto Buizza – Ensemble prediction of COVID-19 – BMJ Open (24 March 2020)*# 11 Appendix A: observed number of COVID-19 in China, Italy South Korea and the United Kingdom

Confirmed cases for China (Table A.1), South Korea (Table A.3) and the United Kingdom (Table A.4) from the World Health Organization web site (https://experience.arcgis.com/). Confirmed cases for Italy (Table A.2), from the Italian Civil Protection Agency web site (http://www.protezionecivile.gov.it/).

| | China | | |
|---|---|---|---|
| Date | Day | Confirmed cases | Cumulative number of confirmed cases |
| 22-Jan | 0.00 | 0.00 | 0.00 |
| 23-Jan | 1.00 | 261 | 261 |
| 24-Jan | 2.00 | 261 | 522 |
| 25-Jan | 3.00 | 462 | 984 |
| 26-Jan | 4.00 | 688 | 1672 |
| 27-Jan | 5.00 | 776 | 2448 |
| 28-Jan | 6.00 | 1800 | 4248 |
| 29-Jan | 7.00 | 1500 | 5748 |
| 30-Jan | 8.00 | 1700 | 7448 |
| 31-Jan | 9.00 | 2000 | 9448 |
| 01-Feb | 10.00 | 2100 | 11548 |
| 02-Feb | 11.00 | 2600 | 14148 |
| 03-Feb | 12.00 | 2800 | 16948 |
| 04-Feb | 13.00 | 3200 | 20148 |
| 05-Feb | 14.00 | 3900 | 24048 |
| 06-Feb | 15.00 | 3700 | 27748 |
| 07-Feb | 16.00 | 3200 | 30948 |
| 08-Feb | 17.00 | 3400 | 34348 |
| 09-Feb | 18.00 | 2700 | 37048 |
| 10-Feb | 19.00 | 3000 | 40048 |
| 11-Feb | 20.00 | 2500 | 42548 |
| 12-Feb | 21.00 | 2000 | 44548 |
| 13-Feb | 22.00 | 3000 (estimated and used instead of the reported value of 15,200; see text) | 47548 |
| 14-Feb | 23.00 | 4000 | 51548 |
| 15-Feb | 24.00 | 2600 | 54148 |
| 16-Feb | 25.00 | 2000 | 56148 |

| 17-Feb | 26.00 | 2100 | 58248 |
|---|---|---|---|
| 18-Feb | 27.00 | 1900 | 60148 |
| 19-Feb | 28.00 | 1800 | 61948 |
| 20-Feb | 29.00 | 396 | 62344 |
| 21-Feb | 30.00 | 892 | 63236 |
| 22-Feb | 31.00 | 825 | 64061 |
| 23-Feb | 32.00 | 649 | 64710 |
| 24-Feb | 33.00 | 221 | 64931 |
| 25-Feb | 34.00 | 517 | 65448 |
| 26-Feb | 35.00 | 412 | 65860 |
| 27-Feb | 36.00 | 439 | 66299 |
| 28-Feb | 37.00 | 329 | 66628 |
| 29-Feb | 38.00 | 435 | 67063 |
| 01-Mar | 39.00 | 574 | 67637 |
| 02-Mar | 40.00 | 206 | 67843 |
| 03-Mar | 41.00 | 129 | 67972 |
| 04-Mar | 42.00 | 119 | 68091 |
| 05-Mar | 43.00 | 143 | 68234 |
| 06-Mar | 44.00 | 145 | 68379 |
| 07-Mar | 45.00 | 103 | 68482 |

*Table A.1. WHO data of confirmed cases for China. Note that in this work, the value reported on day 22 (15,200) was corrected to the mean of days 21 and 23, since it appeared too much of an outlier compared to the others.*

| | Italy | | |
|---|---|---|---|
| **Date** | **Day** | **Confirmed cases** | **Cumulative number of confirmed cases** |
| 21-Feb | 0.00 | 0.00 | 0 |
| 22-Feb | 1.00 | 7 | 7 |
| 23-Feb | 2.00 | 121 | 128 |
| 24-Feb | 3.00 | 101 | 229 |
| 25-Feb | 4.00 | 93 | 322 |
| 26-Feb | 5.00 | 78 | 400 |
| 27-Feb | 6.00 | 128 | 528 |
| 28-Feb | 7.00 | 360 | 888 |
| 29-Feb | 8.00 | 240 | 1128 |
| 01-Mar | 9.00 | 566 | 1694 |
| 02-Mar | 10.00 | 330 | 2024 |
| 03-Mar | 11.00 | 478 | 2502 |
| 04-Mar | 12.00 | 587 | 3089 |

*Roberto Buizza – Ensemble prediction of COVID-19 – BMJ Open (24 March 2020)*

| 05-Mar | 13.00 | 769 | 3858 |
| --- | --- | --- | --- |
| 06-Mar | 14.00 | 778 | 4636 |
| 07-Mar | 15.00 | 1247 | 5883 |
| 08-Mar | 16.00 | 1492 | 7375 |
| 09-Mar | 17.00 | 1797 | 9172 |
| 10-Mar | 18.00 | 977 | 10149 |
| 11-Mar | 19.00 | 2313 | 12462 |
| 12-Mar | 20.00 | 2651 | 15113 |
| 13-Mar | 21.00 | 2547 | 17660 |
| 14-Mar | 22.00 | 3497 | 21157 |
| 15-Mar | 23.00 | 3590 | 24747 |
| 16-Mar | 24.00 | 3233 | 27980 |
| 17-Mar | 25.00 | 3526 | 31506 |
| 18-Mar | 26.00 | 4207 | 35713 |
| 19-Mar | 27.00 | 5322 | 41035 |
| 20-Mar | 28.00 | 5986 | 47021 |
| 21-Mar | 29.00 | 6357 | 53378 |
| 22-Mar | 30.00 | 5760 | 59138 |
| 23-Mar | 31.00 | 4789 | 63927 |
| 24-Mar | 32.00 | 5249 | 69176 |

*Table A.2. Italian Civil Protection data of confirmed cases for Italy.*

| South Korea | | | |
| --- | --- | --- | --- |
| Date | Day | Confirmed cases | Cumulative number of confirmed cases |
| 18-Feb | 0.00 | 0 | 0 |
| 19-Feb | 1.00 | 27 | 27 |
| 20-Feb | 2.00 | 53 | 80 |
| 21-Feb | 3.00 | 98 | 178 |
| 22-Feb | 4.00 | 227 | 405 |
| 23-Feb | 5.00 | 166 | 571 |
| 24-Feb | 6.00 | 231 | 802 |
| 25-Feb | 7.00 | 144 | 946 |
| 26-Feb | 8.00 | 284 | 1230 |
| 27-Feb | 9.00 | 505 | 1735 |
| 28-Feb | 10.00 | 571 | 2306 |
| 29-Feb | 11.00 | 813 | 3119 |
| 01-Mar | 12.00 | 586 | 3705 |
| 02-Mar | 13.00 | 599 | 4304 |
| 03-Mar | 14.00 | 851 | 5155 |
| 04-Mar | 15.00 | 435 | 5590 |
| 05-Mar | 16.00 | 663 | 6253 |
| 06-Mar | 17.00 | 309 | 6562 |
| 07-Mar | 18.00 | 448 | 7010 |
| 08-Mar | 19.00 | 272 | 7282 |
| 09-Mar | 20.00 | 165 | 7447 |
| 10-Mar | 21.00 | 35 | 7482 |
| 11-Mar | 22.00 | 242 | 7724 |
| 12-Mar | 23.00 | 114 | 7838 |
| 13-Mar | 24.00 | 110 | 7948 |
| 14-Mar | 25.00 | 107 | 8055 |
| 15-Mar | 26.00 | 76 | 8131 |
| 16-Mar | 27.00 | 74 | 8205 |
| 17-Mar | 28.00 | 84 | 8289 |
| 18-Mar | 29.00 | 93 | 8382 |
| 19-Mar | 30.00 | 152 | 8534 |
| 20-Mar | 31.00 | 239 | 8773 |
| 21-Mar | 32.00 | 147 | 8920 |
| 22-Mar | 33.00 | 98 | 9018 |
| | | | |
| | | | |

*Table A.3. WHO data of confirmed cases for South Korea.*

| United Kingdom | | | |
|---|---|---|---|
| Date | Day | Confirmed cases | Cumulative number of confirmed cases |
| 26-Feb | 0.00 | 0 | 0 |
| 27-Feb | 1.00 | 3 | 3 |
| 28-Feb | 2.00 | 4 | 7 |
| 29-Feb | 3.00 | 3 | 10 |
| 01-Mar | 4.00 | 13 | 23 |
| 02-Mar | 5.00 | 3 | 26 |
| 03-Mar | 6.00 | 12 | 38 |
| 04-Mar | 7.00 | 36 | 74 |
| 05-Mar | 8.00 | 29 | 103 |
| 06-Mar | 9.00 | 48 | 151 |
| 07-Mar | 10.00 | 45 | 196 |

| 08-Mar | 11.00 | 69 | 265 |
|---|---|---|---|
| 09-Mar | 12.00 | 43 | 308 |
| 10-Mar | 13.00 | 62 | 370 |
| 11-Mar | 14.00 | 77 | 447 |
| 12-Mar | 15.00 | 130 | 577 |
| 13-Mar | 16.00 | 208 | 785 |
| 14-Mar | 17.00 | 342 | 1127 |
| 15-Mar | 18.00 | 251 | 1378 |
| 16-Mar | 19.00 | 152 | 1530 |
| 17-Mar | 20.00 | 407 | 1937 |
| 18-Mar | 21.00 | 676 | 2613 |
| 19-Mar | 22.00 | 643 | 3256 |
| 20-Mar | 23.00 | 714 | 3970 |
| 21-Mar | 24.00 | 1035 | 5005 |
| 22-Mar | 25.00 | 665 | 5670 |
| 23-Mar | 26.00 | 967 | 6637 |

*Table A.4. WHO data of confirmed cases for the United Kingdom.*